\documentclass[ajl]{emulateapj}

\usepackage{graphicx,epsfig,natbib,color}
\usepackage{lscape}
\usepackage{graphicx}
\usepackage{epsfig}
\usepackage{natbib}

\begin{document}
\title{The Driver of Coronal Mass Ejections in the Low Corona: A Flux Rope}

\author{X. Cheng\altaffilmark{1,2,3}, J. Zhang\altaffilmark{2}, M. D. Ding\altaffilmark{1,3},
Y. Liu\altaffilmark{4}, \& W. Poomvises\altaffilmark{2,5,6}}

\affil{$^1$ School of Astronomy and Space Science, Nanjing University, Nanjing 210093, China}\email{xincheng@nju.edu.cn}
\affil{$^2$ School of Physics, Astronomy and Computational Sciences, George Mason University, Fairfax, VA 22030, USA}\email{jzhang7@gmu.edu}
\affil{$^3$ Key Laboratory for Modern Astronomy and Astrophysics (Nanjing University), Ministry of Education, Nanjing 210093, China}
\affil{$^4$ Space Sciences Laboratory, University of California, Berkeley, CA 94720, USA}
\affil{$^5$ Catholic University of America, Washington, DC 20064, USA}
\affil{$^6$ NASA Goddard Space Flight Center, Greenbelt, MD 20771, USA}
\begin{abstract}

Recent \textit{Solar Dynamic Observatory} observations reveal that coronal mass ejections (CMEs) consist of a multi-temperature structure: a hot flux rope and a cool leading front (LF). The flux rope first appears as a twisted hot channel in the Atmospheric Imaging Assembly 94 {\AA} and 131 {\AA} passbands. The twisted hot channel initially lies along the polarity inversion line and then rises and develops into the semi-circular flux rope-like structure during the impulsive acceleration phase of CMEs. In the meantime, the rising hot channel compresses the surrounding magnetic field and plasma, which successively stack into the CME LF. In this paper, we study in detail two well-observed CMEs occurred on 2011 March 7 and 2011 March 8, respectively. Each of them is associated with an M-class flare. Through a kinematic analysis we find that: (1) the hot channel rises earlier than the first appearance of the CME LF and the onset of the associated flare; (2) the speed of the hot channel is always faster than that of the LF, at least in the field of view of AIA. Thus, the hot channel acts as a continuous driver of the CME formation and eruption in the early acceleration phase. Subsequently, the two CMEs in white-light images can be well reproduced by the graduated cylindrical shell flux rope model. These results suggest that the pre-existing flux rope plays a key role in CME initiation and formation.

\end{abstract}
Online-only material: animations, color figures

\section{Introduction}

Coronal mass ejections (CMEs) are large-scale solar eruption activity, releasing a vast amount of plasma and magnetic field into the interplanetary space. They may severely affect the space environment around the Earth \citep[e.g.,][]{gosling93,webb94}. Coronagraph observations show that CMEs generally have a three-part structure: a bright leading front (LF), a dark cavity, and an inner bright core \citep[e.g.,][]{illing83}. The cavity is usually believed to be a helical flux rope \citep[e.g.,][]{gibson06,riley08}. The bright core is thought to be cool filament/prominence matter that is suspended in magnetic dips of a flux rope configuration \citep[e.g.,][]{guo10,jing10}. The flux rope structure has been revealed by Faraday rotation observations that can measure CME magnetic field \citep[e.g.,][]{liuy07}, and recently validated by well-separated multi-spacecraft in situ measurements \citep{liuy08}.

The flux rope structure is also inferred from frequent appearance of sigmoids in EUV and X-ray observations of the solar corona \citep{canfield99,mckenzie08}. These sigmiod structures are composed of two bundles of opposite J-shaped loops, and then form a twisted configuration by magnetic reconnection \citep[e.g.,][]{mckenzie08,liu10,green11}. Many authors successfully simulated the formation of the flux rope. \citet{amari00,amari03} imposed a slow converging motion toward the polarity inversion line (PIL) to the footpoints of field lines to produce the flux rope. Using flux cancellation in a bald-patch separatrix, \citet{aulanier10} transformed the preexisting sheared arcades into a flux rope. \citet{torok03,torok05} and \citet{kliem06} did a theoretical analysis for the flux rope and found that the kink and/or torus instability of the flux rope can initiate a CME. \citet{olmedo10} also found that the eruption of the flux rope can be fully driven by a partial torus instability \citep[see also,][]{williams05}. The flux rope can also be formed in a nonlinear force-free field model based on the observed photospheric vector magnetic fields \citep[e.g.,][]{canou09,canou10,guo10,cheng10b}.

Another key issue in understanding CMEs is the relationship between CMEs and their associated flares. \citet{zhang01,zhang04} studied the kinematics of flare-associated CMEs and proposed a scenario of three phases of CME evolution: a slow initiation phase, an impulsive acceleration phase, and then a propagation phase. Further, they found that the three kinematic phases are closely related to the three phases of the associated flare: the pre-flare phase, flare rise phase, and decay phase, respectively \citep[see also,][]{bur04,cheng10a}. \citet{qiu04} and \citet{temmer08,temmer10} studied the temporal correlation between CME acceleration and the flare hard X-ray (HXR) flux and found that they coincide in time very well. These results suggest that the CMEs are coupled together with the associated flares by the same physical mechanism, especially during the main energy release phase, presumably via magnetic reconnection \citep{lin00,priest02,zhang06}.

Recently, \citet{cheng11} provided possible observational evidence for magnetic reconnection that may form the flux rope fully in the impulsive acceleration phase of a CME. They found that a plasma blob appeared in the Atmospheric Imaging Assembly \citep[AIA;][]{lemen11} passbands of high temperatures (94 {\AA} and 131 {\AA}), rose from the solar limb, then rapidly moved outward, and finally developed into a flux rope-like structure \citep[see also,][]{su12,vourlidas2012}. While in the cooler AIA passbands (171 {\AA} and 211 {\AA}), the background magnetic field and plasma were compressed to form the CME LF. This resulted in that the CME in the inner corona is composed of the multi-temperature structures: the hot flux rope-like channel and the cool LF \citep{cheng11,cheng12b}.

Because the associated source region was partially blocked by the solar limb, \citet{cheng11} was not able to determine the CME structure prior to the eruption. In order to resolve this issue, \citet{zhang12} further studied one CME originating on the solar disc. They found that a twisted channel first appeared in the low corona before the CME eruption and was visible in the hot 131 {\AA} and 94 {\AA} passbands. The twisted hot structure initially lay close to the surface and along the PIL. Once the impulsive acceleration phase began, it erupted explosively outward and developed into the flux rope-like structure. \citet{zhang12} thus concluded that the flux rope, as observed as the hot channel, exists prior to the eruption. In this paper, we investigate the kinematics of two CMEs in detail, especially looking into the relations between the observed hot channel and the LF. We find strong observational evidence that the CME hot channel acts as the engine that drives CME formation and acceleration, at least in the early acceleration phase. In Section 2, we describe the data. In Section 3, we present the observations and results, which are followed by a summary and discussion in Section 4.

\section{Instruments}
AIA on board \textit{SDO} has four telescopes that observe the solar atmosphere almost simultaneously from the photosphere up to the corona using ten narrow UV and EUV passbands. In this paper, we mainly use five passbands including 131 {\AA} (Fe VIII, Fe XXI), 94 {\AA} (Fe XVIII), 211 {\AA} (Fe XIV), 171 {\AA} (Fe IX), and 304 {\AA} (He II); they provide a range of effective temperatures from 0.05 to 20 MK \citep[see][for response functions of AIA channels]{lemen11}. AIA's capability of multi-temperature coverage allows to image different temperature structures of CMEs. During the impulsive phase of a CME, the 131 {\AA} and 94 {\AA} passbands observe the hot plasma structures with a temperature range of $\sim$7--11 MK, the 211 {\AA} and 171 {\AA} passbands primarily observe the CME cavity and LF with a temperature range of $\sim$0.6--3 MK \citep{cheng11,cheng12b,zhang12}, and the 304 {\AA} passband reveals the associated filament material.
Note that, the AIA 131 {\AA} passband also includes the emission from the cool plasma in the transition region in addition to the high temperature plasma from the corona. AIA's temporal cadence, pixel size and field of view (FOV) are 12 seconds, 0.6$''$, and 1.3$R_\odot$, respectively. The multi-temperature capability and super-high cadence provide a unique opportunity to study CME formation in the low corona in detail. Coronagraphs on board $STEREO$ \citep[COR1 and COR2;][]{howard08} and $SOHO$ \citep[LASCO;][]{Brueckner95} image CMEs in white light. In addition, \textit{GOES} X-ray data reveal the temporal profile of the soft X-ray flux for associated flares. The $Reuven$
$Ramaty$ $High$ $Energy$ $Solar$ $Spectroscopic$ $Imager$ \citep[$RHESSI$;][]{lin02} provides the hard X-ray spectrum and imaging of the flares.

\section{Observations and Results}
On 2011 March 7 and 8, two limb solar eruptions took place and were observed simultaneously by \textit{SDO}, \textit{STEREO}, and \textit{SOHO}. Each of these two events was associated with an M-class flare from NOAA active regions (ARs) 11164 and 11171, respectively. AIA 94 {\AA} and 131 {\AA} ($\sim$7--11 MK) images show that a twisted hot channel appeared along the PIL before the onset of the flare. \citet{zhang12} presented the evolution of the flux rope-like hot channel for 2011 March 8 event. Here we quantitatively analyze the full evolution of the CME hot channel and LF for both events with much more details.

\subsection{Initiation Phase of CME Hot Channels}
\begin{figure} 
     \vspace{-0.0\textwidth}    
     \centerline{\hspace*{0.00\textwidth}
               \includegraphics[width=1.\textwidth,clip=]{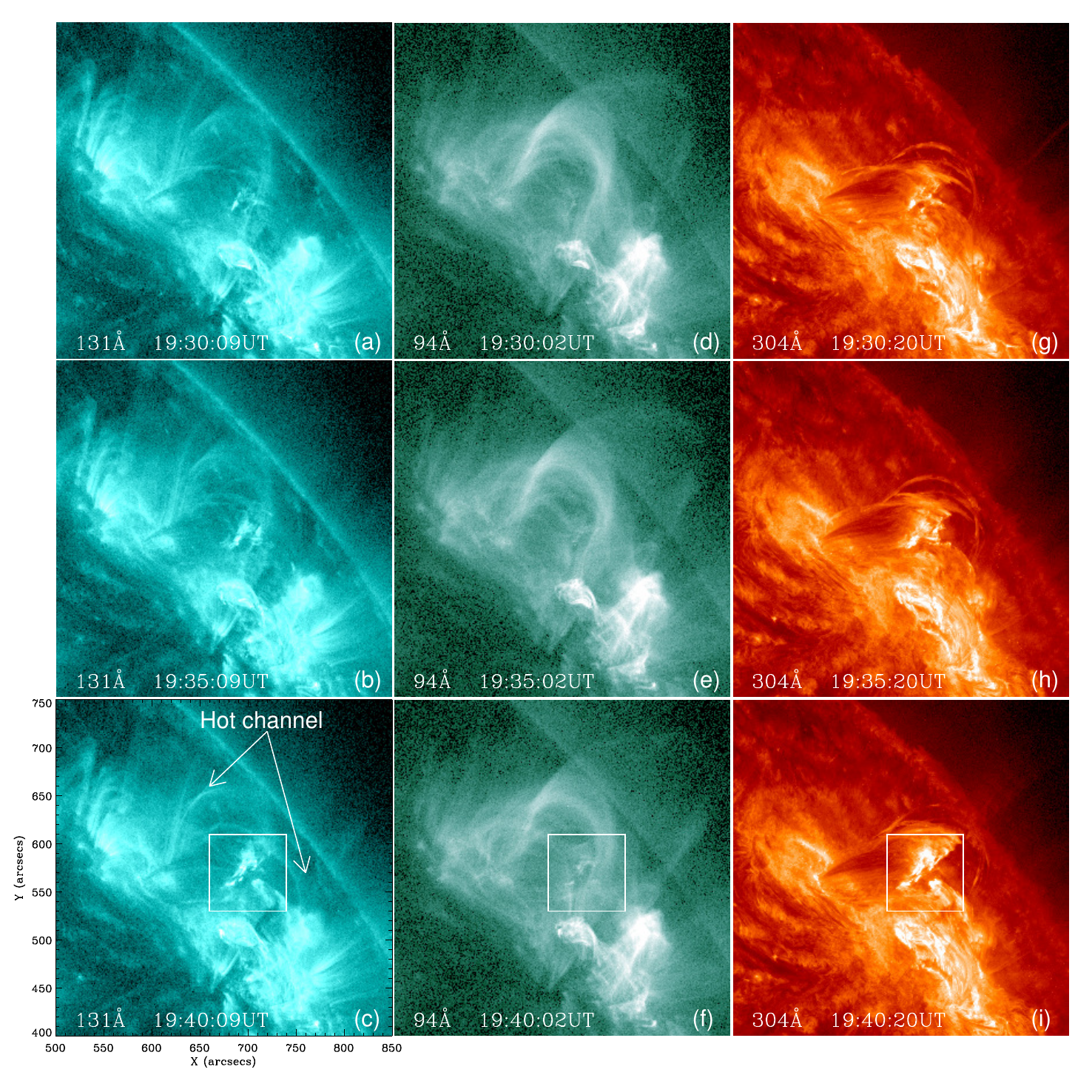}
               }
\vspace{0.0\textwidth}   
\caption{(a--f) AIA 131 {\AA} ($\sim$0.4 MK and 11 MK) and 94 {\AA} ($\sim$7 MK) images of the 2011 March 07 CME hot channel in its initiation phase; (g--i) 304 {\AA} ($\sim$0.05 MK) images of the associated filament. The hot channel is pointed out by two white arrows in panel (c), the brightening is indicated by the small box in panel (c), (f), and (i).} \label{f1}

(Animations of this figure are available in the online journal.)
\end{figure}

\begin{figure} 
     \vspace{-0.0\textwidth}    
     \centerline{\hspace*{0.00\textwidth}
               \includegraphics[width=1.\textwidth,clip=]{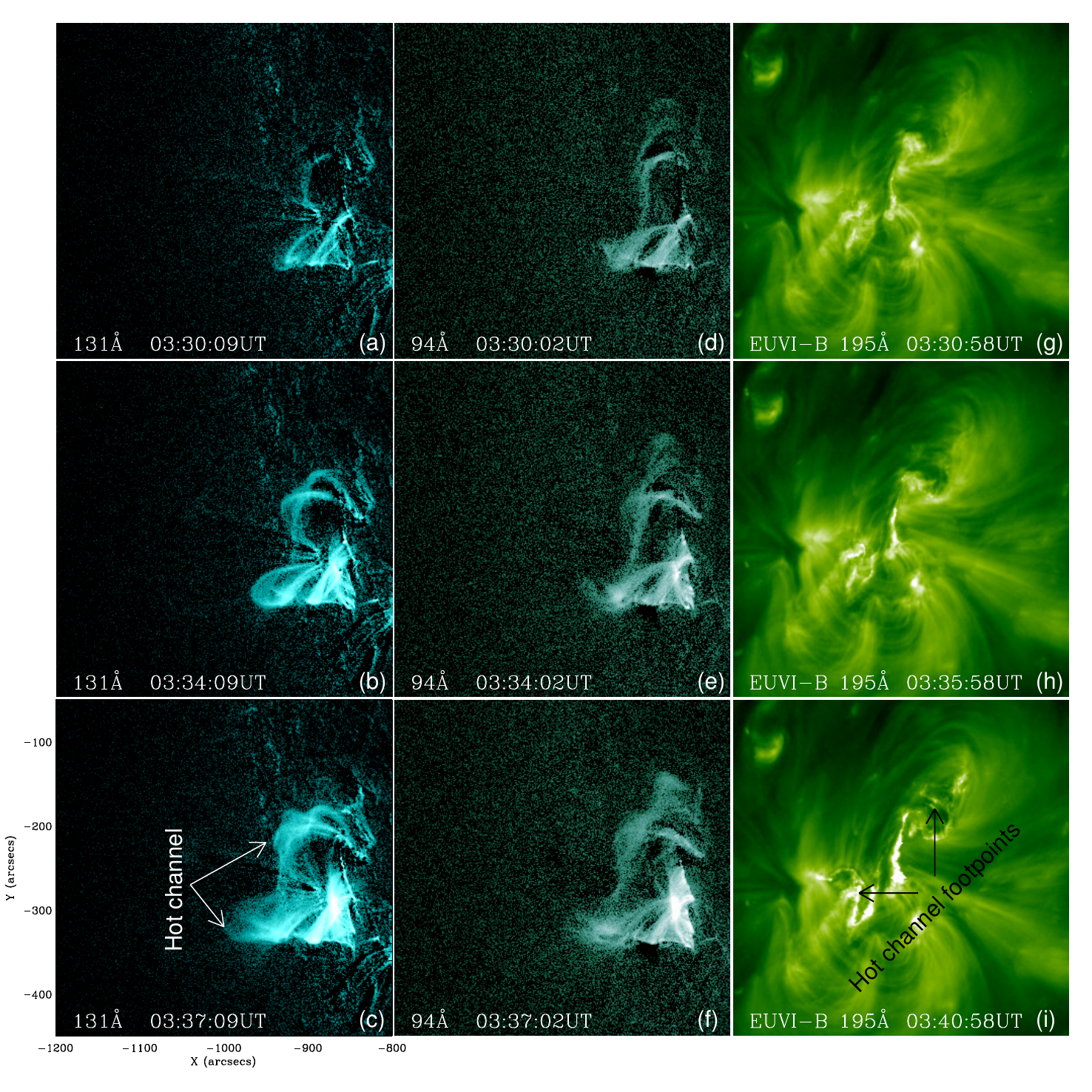}
               }
\vspace{0.0\textwidth}   
\caption{(a--f) AIA 131 {\AA} ($\sim$0.4 MK and 11 MK) and 94 {\AA} ($\sim$7 MK) base-difference images of the 2011 March 08 CME hot channel in its initiation phase; (g--i) EUVI-B 195 {\AA} images of the associated source region. The hot channel and its footpoints in the source region are indicated by two white and black arrows, respectively.} \label{f2}

(An animation of this figure is available in the online journal.)
\end{figure}

The AIA 94 {\AA} and 131 {\AA} images of the hot channels of the 2011 March 7 and 8 CMEs are shown in Figures \ref{f1} and \ref{f2}. For both events, a hot channel appeared prior to the onset of corresponding flare by several minutes.

\begin{figure} 
     \vspace{-0.0\textwidth}    
     \centerline{\hspace*{0.00\textwidth}
               \includegraphics[width=0.48\textwidth,clip=]{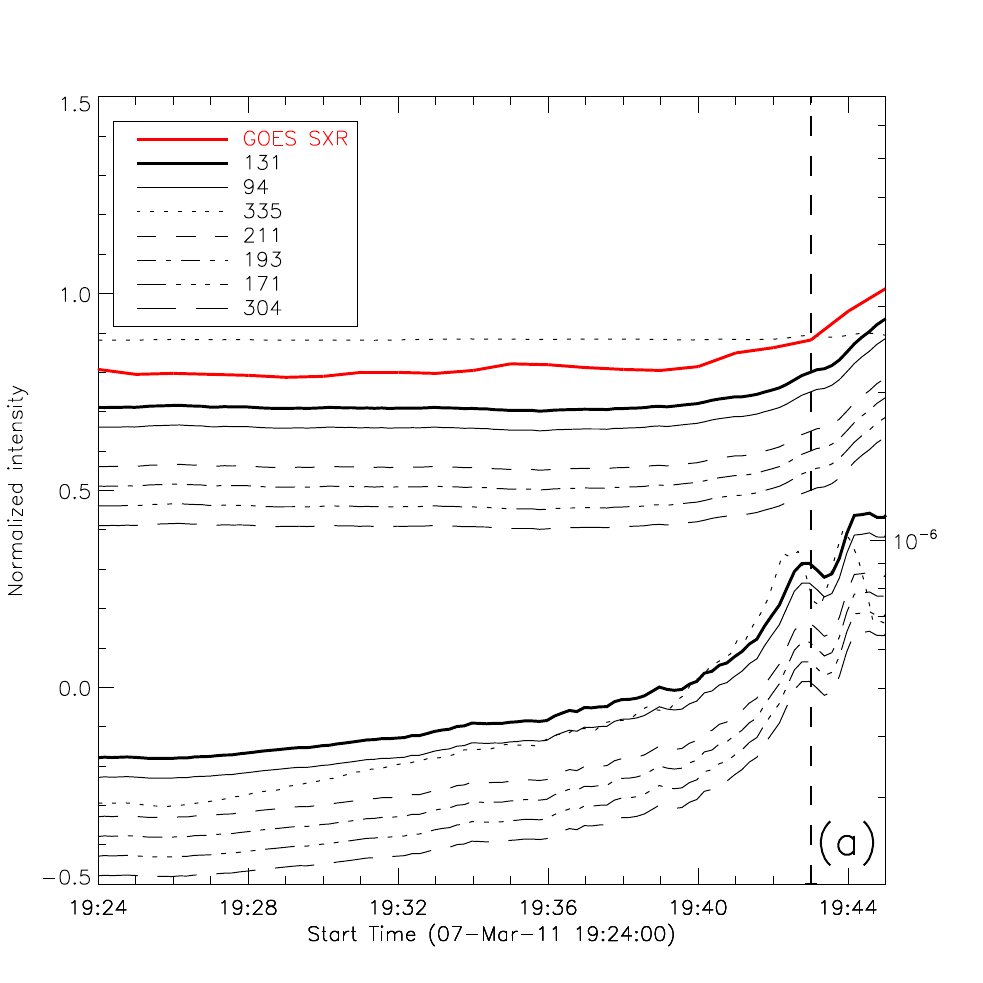}
               \includegraphics[width=0.48\textwidth,clip=]{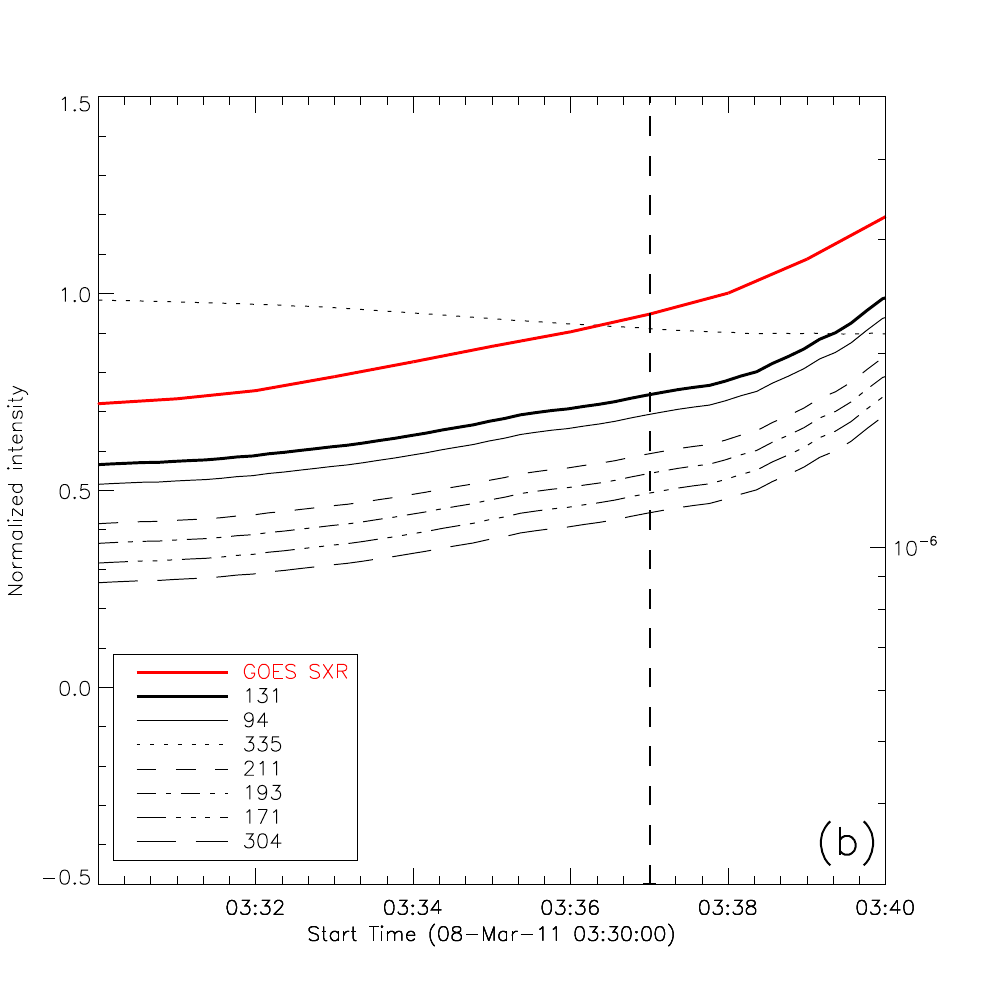}
               }
\vspace{0.0\textwidth}   
\caption{(a) $GOES$ soft X-ray 1--8 {\AA} flux and EUV integral intensity at 131 {\AA} ($\sim$0.4 MK and 11 MK), 94 {\AA} ($\sim$7 MK), 335 {\AA} ($\sim$2.5 MK), 211 {\AA} ($\sim$2 MK), 193 {\AA} ($\sim$1.3 MK and 20 MK), 171 {\AA} ($\sim$0.6 MK), and 304 {\AA} ($\sim$0.05 MK). The bundle of upper and bottom lines refer to the FOV of Figure 1 and the box, respectively. (b) Same as (a) but for the FOV of Figure 2. These EUV intensities are normalized to their maxima and shifted downward to avoid overlaying. The vertical dotted lines show the onset time of the flares in $GOES$ observations.} \label{f3}
\end{figure}

As for the 2011 March 7 event, an interesting feature is that the cool filament material appeared along the magnetic structure of the hot channel, which is identified in the 304 {\AA} passband (Figure \ref{f1}(g--i)). From the 304 {\AA} images, one also notices a brightening that started to appear after $\sim$19:20 UT. The brightening occurred inside the filament, which indicates that magnetic reconnection initially took place there. The reconnection gradually heated the plasma in the filament and made the intensity in 304 {\AA} passband increase. With the heating ongoing, the brightening was further enhanced and extended upward. At the same time, the whole filament structure started to rise slowly. By inspecting visually the movies of the 131 {\AA} and 304 {\AA} passbands (available in the online version), we infer possible existence of an X-type reconnection point inside the filament structure, and the location of the brightening is almost coincident with that of the X-type point, which is shown in the small box of Figure \ref{f1}(c) and (f). In order to compare the emission produced at the X-type point with the $GOES$ soft X-ray 1--8~\AA\ flux, we plot the temporal profiles of the EUV intensity for all AIA EUV passbands (Figure \ref{f3}(a)), integrated over the brightening region shown by the small box in Figure \ref{f1}. The temporal profiles of the intensity in the whole active region are also plotted. One can see that the intensity of the brightening in all AIA EUV passbands rapidly increase about 10 minutes prior to the beginning of the associated flare ($\sim$19:43 UT). However, for the whole AR no apparent increase of the emission intensity appears, the evolution is basically consistent with that of the $GOES$ soft X-ray 1--8~\AA\ flux.

Similarly, for the 2011 March 8 CME, the hot channel appearing in the 131 {\AA} and 94 {\AA} passbands is shown in Figure \ref{f2}(a--f). It became first visible at $\sim$03:30 UT (Figure \ref{f2}(a) and (d)). With time elapsing, it took on a twisted shape with a double elbow above at $\sim$03:34 UT. Then, the twisted hot channel rose slowly with the southern elbow field slipping down (Figure \ref{f2}(b) and (e)). Just prior to the onset of the associated flare, the hot channel apparently presented a twisted shape and had fully detached from the elbow field lines (Figure \ref{f2}(c) and (f)). By checking the EUVI-B 195~\AA\ images, we find that the 2011 March 8 event originated from a forward sigmoid active region (Figure \ref{f2}(g)--(i)). The hot channel lay along the PIL, which can be indicated by the connection line of the two footpoints of the channel (black arrows in Figure \ref{f2}(i)). It is worth noting that, for this case, there is no filament material associated with the hot channel structure. This may be the reason why the hot channel is only visible in the 131 {\AA} and 94 {\AA} passbands but not the 304 {\AA} passband. We also compare the temporal evolution of the integrated flux intensity of the whole active region with the $GOES$ soft X-ray 1--8~\AA\ flux. From Figure \ref{f3}(b) we see that, except the 335 {\AA} emission, the AIA EUV intensity has started to increase slightly along with the $GOES$ soft X-ray flux prior to the beginning of the associated flare ($\sim$03:37 UT). The decrease of the intensity in 335 {\AA} passband is probably caused by the heating moving the plasma out of the temperature response range sensitive to this passband.

\begin{figure} 
     \vspace{-0.0\textwidth}    
     \centerline{\hspace*{0.08\textwidth}
               \includegraphics[width=0.44\textwidth,clip=]{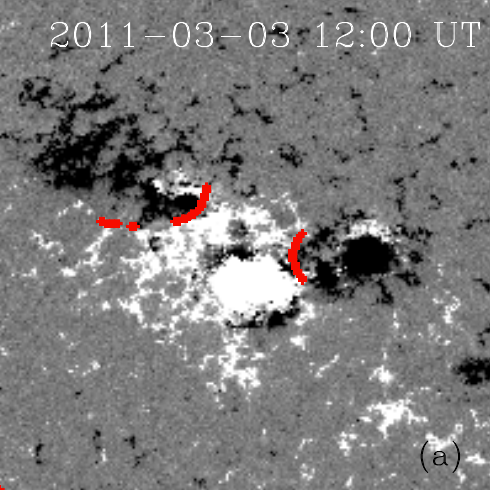}
               \includegraphics[width=0.44\textwidth,clip=]{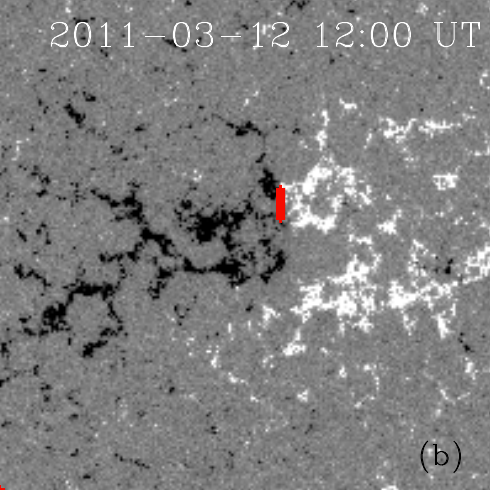}
               }
     \centerline{\hspace*{0.00\textwidth}
               \includegraphics[width=1.0\textwidth,clip=]{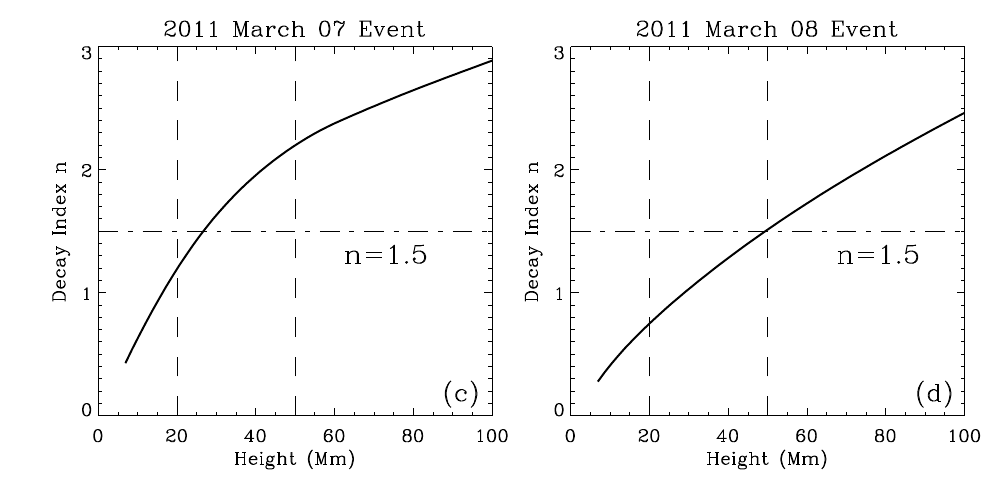}
               }

\vspace{0.0\textwidth}   
\caption{(a) Line-of-sight magnetogram at 12:00 UT on 2011 March 3 observed by \emph{SDO}/HMI with the main neutral lines (red lines) overlaid. (b) Same as (a) but on 2011 March 12. (c) and (d) Distributions of the decay index of the background magnetic field with height, dash-dotted line shows the decay index of 1.5, two dashed lines point out the heights of 20 and 50 Mm above the photosphere.}  \label{f4}
\end{figure}

Here, we also study the property of the magnetic field associated with the eruptions. It is generally accepted that the decay index of the background magnetic field ($n=-d \ln B/d \ln h$; B denotes the background magnetic field strength at the
geometrical height h above the solar surface) controls the torus instability of the flux rope; the torus instability will take place if the decay index $n$ is greater than a threshold value \citep{torok05,kliem06,olmedo10,fan10,cheng10a}. Here, the decay index of the background field is calculated from the potential field model based on the line-of-sight magnetograms observed by the Helioseismic and Magnetic Imager \citep[HMI;][]{schou12} onboard \emph{SDO} (Figure \ref{f4}(a) and (b)). Note that, in calculating the decay index, only the transverse component of coronal magnetic field is used, since the line-of-sight component does not contribute to the downward constraining force exerting onto the flux rope. The final decay index is an average value over the main PILs (red lines) of AR. The decay index distributions with the height for 2011 March 7 and 8 events are displayed in Figure \ref{f4}(c) and (d), respectively. Through the AIA 131 {\AA} images, we find that the hot channels initially located at the height of $\sim$ 20 Mm over the AR (see Figure \ref{f7}(a), \ref{f8}(a), \ref{f9}(b), and \ref{f10}(b)), where the decay indexes are always smaller than 1.5 \citep[a nominal critical value of the torus instability;][]{torok05}, thus may not be able to trigger the torus instability. Taking the pre-flare brightening into account, we argue that it is most likely that the magnetic reconnection drives the slow rise phase of the two hot channels. This reconnection during the initiation phase should be similar to the suggested first step reconnection in a tether-cutting model which helps the formation and rise of the flux rope \citep{moore01,aulanier10}.

Furthermore, we study the transition of the hot channel from the slow rise phase to the impulsive acceleration phase. At the end of the slow rise phase, the 2011 March 7 (8) hot channel obtained the velocity of $\sim$200 km s$^{-1}$ ($\sim$100 km s$^{-1}$) and attained the height of $\sim$50 Mm (indicated by the arrows in Figure \ref{f7}(a) and \ref{f8}(a) and the vertical dashed lines in Figure \ref{f9}(b) and \ref{f10}(b)), where the decay index of the background field is greater or only slightly greater than 1.5 (Figure \ref{f4}(c) and (d)), thus possibly triggering the torus instability and playing an important role in the transition from the slow rise phase to the impulsive acceleration phase. For the 2011 March 8 hot channel, the transition occurred exactly at the height of $\sim$50 Mm; while for the 2011 March 7 one, the transition possibly took place at the lower height, such as at $\sim$30 Mm, where the decay index has been larger than the critical value 1.5. Moreover, we note that the time of the magnetograms we used is several days before or after that of the CME eruption and the exact location of the hot channel in the AR is unknown, which may have some influence on the calculation of the decay index and the judgment of the torus instability occurrence.

\subsection{Kinematics of CME Flux Rope and Leading Front}

\begin{figure} 
     \vspace{-0.0\textwidth}    
     \centerline{\hspace*{0.00\textwidth}
               \includegraphics[width=1.\textwidth,clip=]{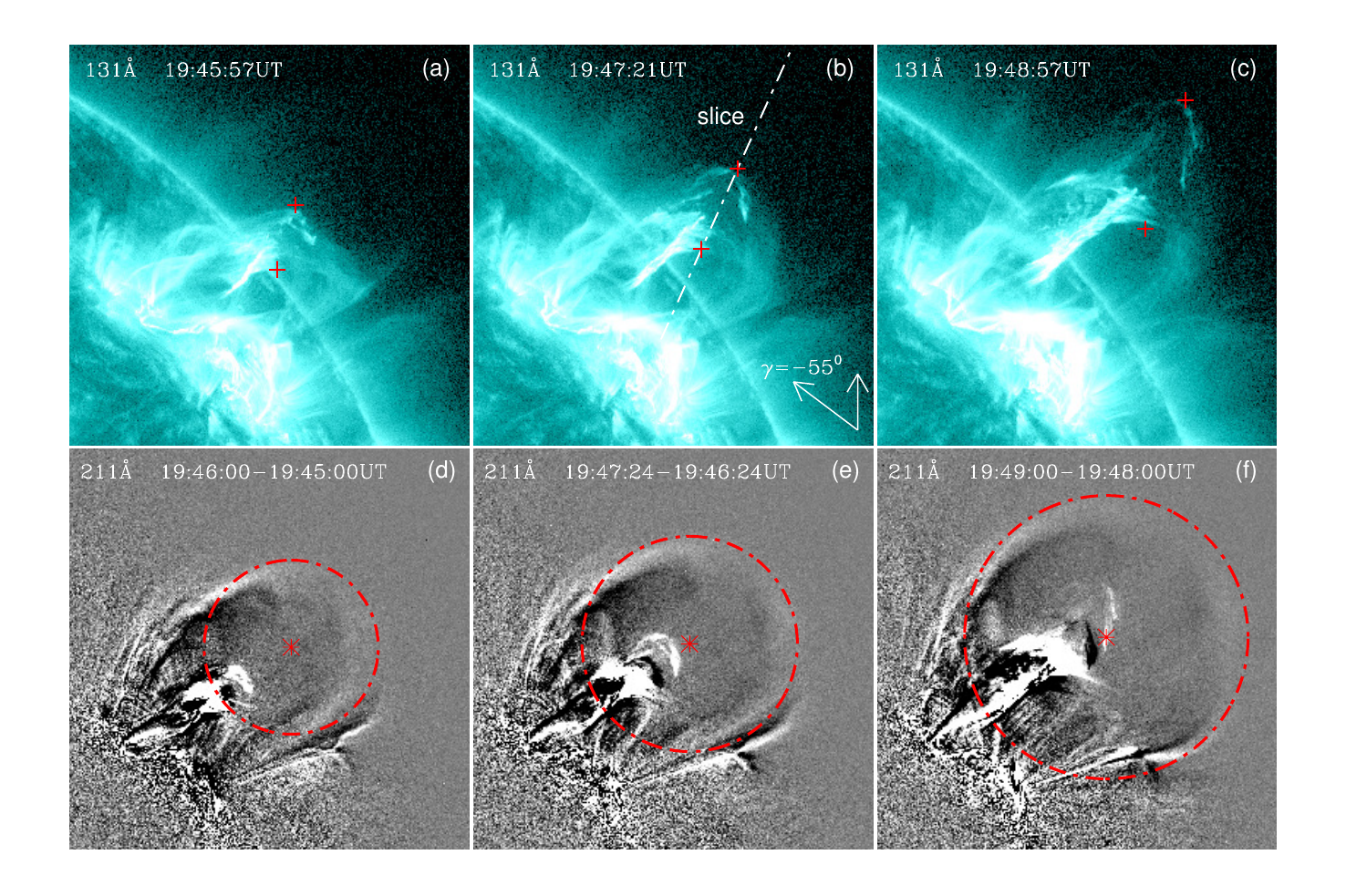}
               }
\vspace{0.0\textwidth}   
\caption{(a--c) AIA 131 {\AA} ($\sim$0.4 MK and 11 MK) images of the 2011 March 07 CME hot channel in the acceleration phase. Two pluses in each panel show the top and bottom positions of the hot channel. In panel b, the dashed line indicates the slice orientation, the vertical and inclined arrows show the north pole and the direction of the hot channel axis, respectively. (d--f) Circular fitting of 2011 March 07 CME bubble. The crosses mark the center of the circle. The bottom panels (d-f) are difference images at 211 {\AA} between the times shown in each panel.} \label{f5}
\end{figure}

\begin{figure} 
     \vspace{-0.0\textwidth}    
     \centerline{\hspace*{0.00\textwidth}
               \includegraphics[width=1.\textwidth,clip=]{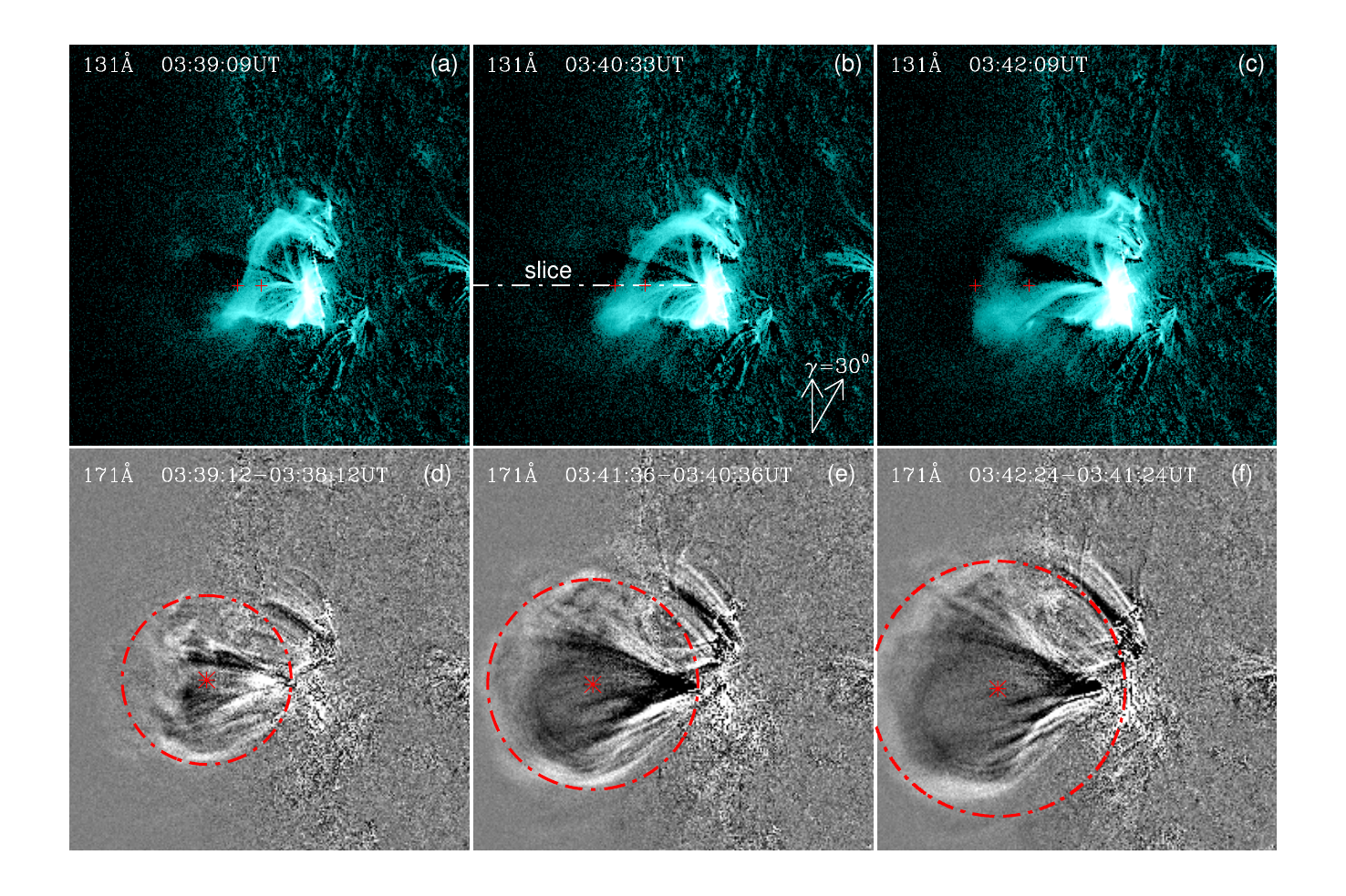}
               }
\vspace{0.0\textwidth}   
\caption{(a--c) AIA 131 {\AA} ($\sim$0.4 MK and 11 MK) images of the 2011 March 08 CME hot channel in the acceleration phase. The two pluses in each panel show the top and bottom positions of the hot channel. In panel b, the dashed line indicates the slice orientation, the vertical and inclined arrows show the north pole and the direction of the hot channel axis, respectively. (d--f) Circular fitting of the 2011 March 08 CME bubble. The crosses show the center of the circle. The bottom panels (d-f) are difference images at 171 {\AA} between the times shown in each panel.}
\label{f6}
\end{figure}

\begin{figure} 
     \vspace{-0.0\textwidth}    
     \centerline{\hspace*{0.00\textwidth}
               \includegraphics[width=0.6\textwidth,clip=]{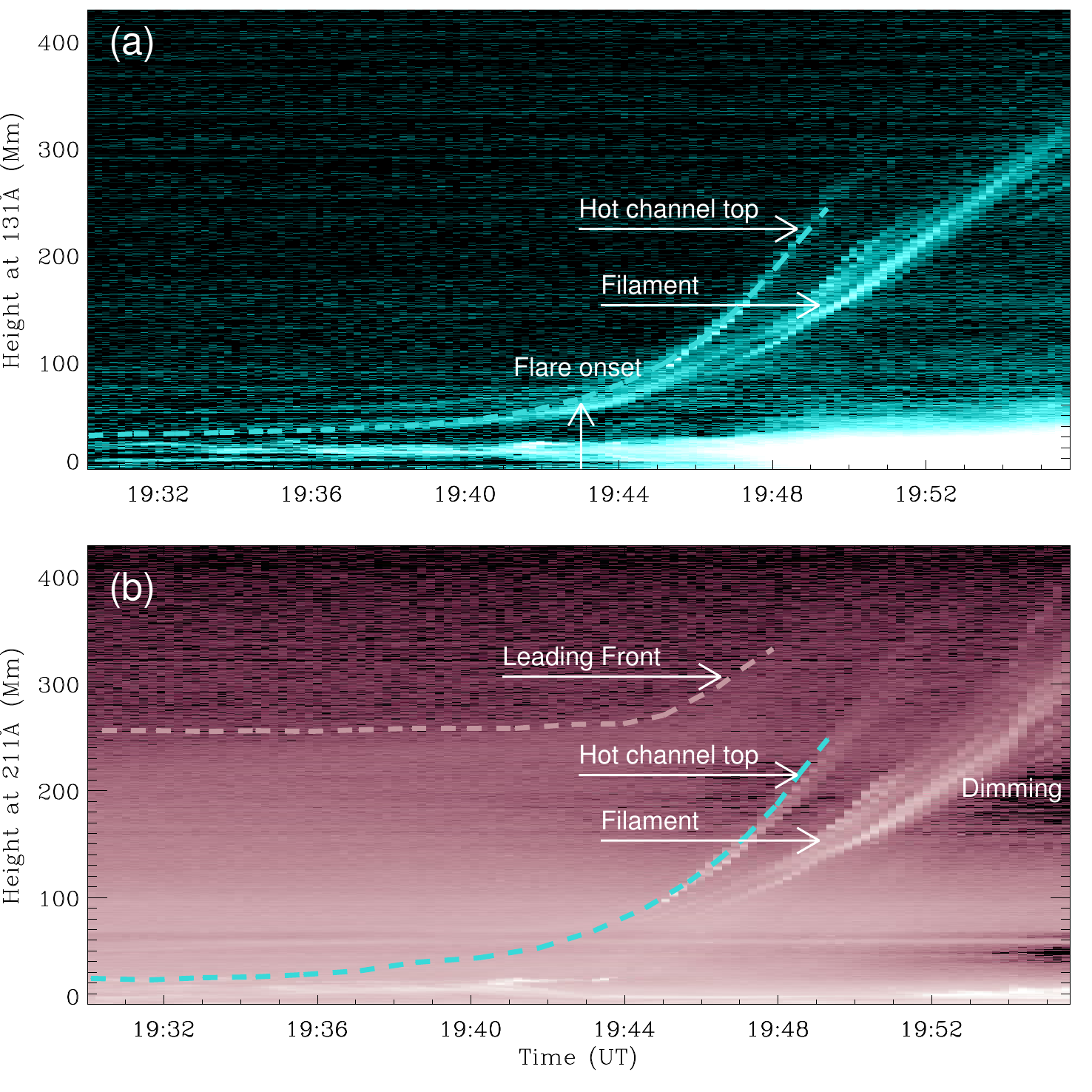}
               }
\vspace{0.0\textwidth}   
\caption{Temporal evolution of the brightness along the slice as shown in Figure \ref{f5}(b) for 131 {\AA} ($\sim$0.4 MK and 11 MK; upper) and 171 {\AA} ($\sim$0.6 MK; lower) base-difference images. An overall picture of the early CME evolution can be seen in a lower temperature passband, as shown in the bottom panel. The vertical arrow indicates the onset of the flare rise phase.} \label{f7}
\end{figure}

When the associated flare starts to brighten in soft X-ray or hard X-ray, a CME often impulsively accelerates outward. For the two CMEs on 2011 March 7 and 8, the CME hot channel started to rapidly rise and expand immediately at the time of the flare onset (Figure \ref{f5}(a)--(c) and Figure \ref{f6}(a)--(c)). In order to clearly present the kinematic evolution of the CME hot channel, we take slices from the images along a direction of the rising motion (Figure \ref{f5}(b) and Figure \ref{f6}(b)) and show the time evolution of the hot channel brightness along these slices (Figure \ref{f7}(a) and Figure \ref{f8}(a)). For the 2011 March 7 event, the hot channel quickly expanded and seemed to split into two parts: the hot channel top and bottom. The bottom part is actually related to the cool filament. The quick rise and expansion of the hot channel also occurred in the 2011 March 8 event without, however, the splitting of the structure (Figure \ref{f8}(a)). From Figure \ref{f5}(d)--(f) and Figure \ref{f6}(d)--(f), we find that the overlying field of the hot channel gradually formed an EUV cavity resembling a bubble. The temporal evolution of the bubble brightness along the same slices is shown in Figure \ref{f7}(b) and Figure \ref{f8}(b). One can see that the CME bubble experienced internal expanding and also external compressing of the surrounding field and plasma to produce the CME LF with enhanced plasma density, especially for the 2011 March 8 event. Note that, in the two cases we do not find any evident features of the EUV shocks, which usually is driven by the early expansion of the CME bubble and appears ahead of the CME LF with a small standoff distance \citep[e.g.,][]{cheng12a,dai12,downs12,pat12,olmedo12}.

By tracking the hot channel in the 131 {\AA} images, we measure the height-time variations of the hot channel top and bottom, as shown by the crosses in Figure \ref{f5}(a)--(c) and Figure \ref{f6}(a)--(c). The distance between the hot channel top and bottom is considered as its width. In order to obtain the height-time measurements of the CME LF, we fit the CME bubble as a circle and take the height of the bubble top as the height of the LF. The radius of the circle is regarded as the size of the CME bubble. The fitting circles are shown in Figure \ref{f5}(d)--(f) and Figure \ref{f6}(d)--(f), in which the bubble center is indicated by a cross. One can find that the circle fits the upper part of the bubble very well in the FOV of AIA. Note that all heights are measured from the surface of the Sun.

We first study the expansion properties of CME structures through the measurement of aspect ratios. The aspect ratio of the hot channel is the ratio between its top height and width, while that of the bubble is the ratio between the height of its center and its radius. The temporal evolution of these aspect ratios are shown in Figure \ref{f9}(a) and Figure \ref{f10}(a). For the 2011 March 7 CME, the aspect ratio of the hot channel kept almost a constant value for $\sim$4 minutes after the flare onset at $\sim$19:43 UT and then decreased from $\sim$4.0 to 3.0 in the next $\sim$3 minutes. The aspect ratio of the bubble changed more rapidly. It continuously decreased from $\sim$2.2 to 1.2 during the 6-minute of measurement. While for the 2011 March 8 CME, the aspect ratio of the bubble kept a constant value of $\sim$1.1 with a slight decrease after $\sim$3 minutes of the onset, the aspect ratio of the hot channel monotonically deceased from $\sim$5.0 to 3.0 during the 6-minute of measurement. These results indicate that the growth rate of the width of the hot channel is generally larger than that of its height during the initial rise phase of the flares. It implies that the hot channel always undergoes an overexpansion (here an overexpansion is defined such that the expansion speed is larger than the bulk propagation speed). For the CME bubble, the evolution is different from case to case. Apparently, for the 2011 March 7 CME, the bubble undergoes an overexpansion; while for the 2011 March 8 event, it mainly has a self-similar expansion. However, note that, the earlier overexpansion of the 2011 March 8 CME bubble than $\sim$03:39 UT may be missed because of the indiscernibility of the initial CME bubble.

Furthermore, we study the kinematics of CME structures. The temporal variations of the height and width of the CME hot channel and bubble are plotted in Figure \ref{f9}(b) and Figure \ref{f10}(b), from which one can see that the CME hot channel is catching up with the LF in the FOV of AIA. Specifically, the distance between the hot channel top and the CME LF decreases. In order to obtain the temporal variations of their speed and acceleration, the height measurements usually need to be smoothed first in order to reduce the uncertainty in the measurement. Here we use a cubic spline smoothing method to reduce uncertainty, and then calculate the speed with a piece-wise numerical derivative method, i.e., the Lagrangian interpolation of three adjacent points \citep[e.g.,][]{zhang01,zhang04,cheng10a,pat10b}. The deduced speed-time profiles are plotted in Figure \ref{f9}(c) and Figure \ref{f10}(c). Note that the uncertainty in the speed mainly results from the error in the height measurements, which is estimated to be 2 pixels for AIA observations. With the same method the CME acceleration can be further derived from the speed, as shown in Figure \ref{f9}(d) and Figure \ref{f10}(d).

For both events, the speed of the CME hot channel is larger than that of the CME LF during the early acceleration phase. It is likely that the hot channel drives the CME bubble upward. For the 2011 March 7 event, the hot channel ascended with an average speed of $\sim$100 km s$^{-1}$ prior to the flare beginning ($\sim$19:43 UT). After the onset of the flare, the speed quickly increased from $\sim$200 km s$^{-1}$ to $\sim$800 km s$^{-1}$ within $\sim$6 minutes. At all times, the speed of the hot channel was $\sim$100 km s$^{-1}$ greater than that of the LF. The speed evolutions of the CME hot channel and LF are synchronized in time with the $GOES$ 1--8~\AA\ soft X-ray flux profile (Figure \ref{f9}(c)). The results also hold for the 2011 March 8 event. The hot channel rose slowly with an average speed of $\sim$60 km s$^{-1}$ in the initiation phase. Subsequently, at each instant the speed of the hot channel was $\sim$50 km s$^{-1}$ greater than that of the LF (Figure \ref{f10}(c)).

In addition, we find that the onset of the acceleration of the hot channels is always earlier than the onset of the hard X-ray flux of the flares, as well as the first appearance of the LF. The acceleration evolutions of the CME hot channel and LF, as well as the RHESSI 15--25 keV hard X-ray curves of the associated flares, are displayed in Figure \ref{f9}(d) and Figure \ref{f10}(d). One can see that the hot channel on 2011 March 7 had begun to accelerate at $\sim$19:41 UT while both the onset of the 15--25 keV hard X-ray flux of associated flare and the first appearance of the LF at later $\sim$19:43 UT. For the 2011 March 8 CME, the acceleration of the hot channel took place at $\sim$03:35 UT, whereas the 15--25 keV hard X-ray flux of associated flare begins at $\sim$03:37 UT and the LF appears first at $\sim$03:38 UT. The leading time is $\sim$2 minutes for both events. The results imply that the CME hot channel plays an important role in inducing the energy releases of flares and in forming the CME LF, also imply that the transition of the hot channel from the slow rise to the impulsive acceleration probably take place earlier than the flare onset. Nevertheless, due to that it is very difficult to determine the exact onset of the impulsive acceleration caused by the torus instability, we thus use the flare onset as an approximation of the transition onset in previous Section.

The speed evolution of the CME hot channel and the LF is generally consistent in time with the variation of the soft X-ray flux although the hot channel acts as a driver in the CME formation and eruption. After the CMEs have left the FOV of AIA, they are well observed by COR1, COR2 and LASCO. Since the two events are close to the solar limb, the height-time measurements from LASCO are only slightly affected by the projection effect. As for COR1 and COR2 observations, we use the ``SCC\b{ }MEASURE" routine in the SSW package to estimate the three-dimension heights \citep[e.g.,][]{li11}. All the height-time data are plotted in Figure \ref{f9}(e) and Figure \ref{f10}(e) including the measurements from AIA. The corresponding speed is plotted in Figure \ref{f9}(f) and Figure \ref{f10}(f). For the 2011 March 7 CME, the speed profile of the CME is coincident in time very well with the variation of the $GOES$ soft X-ray 1--8~\AA\ flux. The CME speed increases to $\sim$2000 km s$^{-1}$ at $\sim$20:10 UT and then decreases to $\sim$1500 km s$^{-1}$ at $\sim$21:10 UT. The 2011 March 8 CME seems to keep a constant speed of $\sim$1100 km s$^{-1}$ after the peak time of the flare. Due to the lack of high-cadence measurements during the later rise phase of the associated flare, we can not determine exactly (i.e., in one minute of accuracy) when the CME get its maximum speed. Note that for COR1, COR2, and LASCO observations, the uncertainty of measuring the LF height is estimated as 4 pixels.

\subsection{Flux Rope Fitting of CMEs in White Light Images}
The graduated cylindrical shell (GCS) model is proposed by \citet{thernisien06} and represents CMEs as flux rope-like structures. It consists of a tubular section forming the main body and two cones rendering the ``legs" of the CME. The geometry is shown in Figure 1 of \citet{thernisien06,thernisien09}. The model is controlled by six free parameters: Carrington longitude ($\phi$) and latitude ($\theta$) of the source region, the tilt angle ($\gamma$) of the flux rope, the height ($r$) of the CME LF, the half-angle ($\alpha$) between the two legs of the flux rope, and the aspect ratio ($\kappa$) of the CME flux rope. Note that $\kappa$ is defined by \citet{thernisien06} as the ratio between the minor radius ($\omega$) of the flux rope and the height ($r$). It is different from the definitions of aspect ratios of the CME hot channel and bubble in this paper. But, the two kinds of definitions can be converted to each other \citep[see][page 9]{pat10a}. The GCS model has been implemented in many aspects, e.g., fitting the CME cavity \citep{pat10a}, determining flux rope orientation and comparing the results with in situ measurements \citep{liuy10}, and studying the kinematics and expansion of CMEs \citep{yod10}. In this section, we test whether it can reproduce the morphology of the 2011 March 7 and 8 CMEs and then determine the orientation of the flux ropes in white-light observations, comparing with that in EUV observations.

Using the EUVI 195~\AA\ images, we first estimate $\phi$ and $\theta$ through the location of the AR. Then we vary $\alpha$, $\kappa$, $\gamma$, and $r$ until the best approximations to COR2 and C2 images are obtained. In fact, $\alpha$ is usually kept as a constant value; $\kappa$ and $\gamma$ only change slightly when the CME is propagating in the FOV of COR2. The final positioning and model parameters of the two CME flux ropes are listed in Table \ref{tb}. A wireframe rendering is shown in Figures \ref{f11} and \ref{f12}. One can see that the flux rope model reconstructs the CME images very well. For the 2011 March 7 CME, the flux rope propagated toward the northwest in the FOVs of STEREO-B and LASCO, and toward the northeast in the FOV of STEREO-A. In the meantime, it expanded self-similarly ($\kappa$=0.4). At 21:08 UT, it had appeared as a partial halo CME, which was possibly preceded by a shock (Figure \ref{f11}(d)--(f)). The 2011 March 8 CME also expanded self-similarly ($\kappa$=0.3) and may also have caused a shock surrounding the CME (Figure \ref{f12}(d) and (f)), but it had a propagation direction toward the southwest in the FOV of STEREO-B and toward southeast in the FOVs of LASCO and STEREO-A (Figure \ref{f12}). Moreover, by inspecting the orientation of the flux rope, we find that the 2011 March 7 (8) CME have a counter-clockwise (clockwise) rotation while rising. Assuming that the initial orientation of the CME is the orientation of the axis of the hot channel, as shown in Figures \ref{f5}(b) and \ref{f6}(b), the rotation angle is $\sim$--30$^{\circ}$ (25$^{\circ}$) when out of the FOV of COR2. This result is well compatible with the hemispheric preference of the CME rotation \citep[counter-clockwise (clockwise) rotation in the northern (southern) hemisphere;][]{rust96,green07}.

\section{Summary and Discussions}

In this paper, we investigate in detail the formation and evolution of different structural components of two well-observed CMEs. In the initiation phase, the pre-eruption structure appears as a twisted hot channel, having an orientation consistent with that of the PIL and showing a shear with the overlying field. The twisted channel is most likely heated by the magnetic reconnection during the initiation phase, making it visible in the high temperature passbands, e.g., 131 {\AA} and 94 {\AA}. The reconnection location may be inside or around the hot channel and the corresponding reconnection rate is relatively small, being different from the impulsive phase reconnection underneath the CME flux rope \citep[e.g.,][]{fan10}. As for the physical mechanism of the slow rise phase of the flux rope-like hot channel, we examine the decay index of the background field with height and find that the decay index is small enough at the initial height of the hot channel ($\sim$20 Mm), which possibly does not allow the occurrence of the torus instability. We thus speculate that the mechanism of driving the slow rise of the hot channel is most likely to be magnetic reconnection, but not the one producing X-ray flares. Such reconnection should be similar to the coronal slip-running reconnection in the simulation of the flux rope \citep{aulanier10}. Nevertheless, with the reconnection ongoing, the hot channel rises gradually with a speed of 50--100 km s$^{-1}$ and reach a critical height finally, such as $\sim$50 Mm for the two CMEs studied in this paper, where the background field declines rapidly enough, thus being able to trigger the torus instability, resulting in the onset of the hot channel impulsive acceleration phase and the flare rise phase; such a scenario has been proposed in earlier numerical simulations by \citet{aulanier10}.

\begin{figure} 
     \vspace{-0.0\textwidth}    
     \centerline{\hspace*{0.00\textwidth}
               \includegraphics[width=0.6\textwidth,clip=]{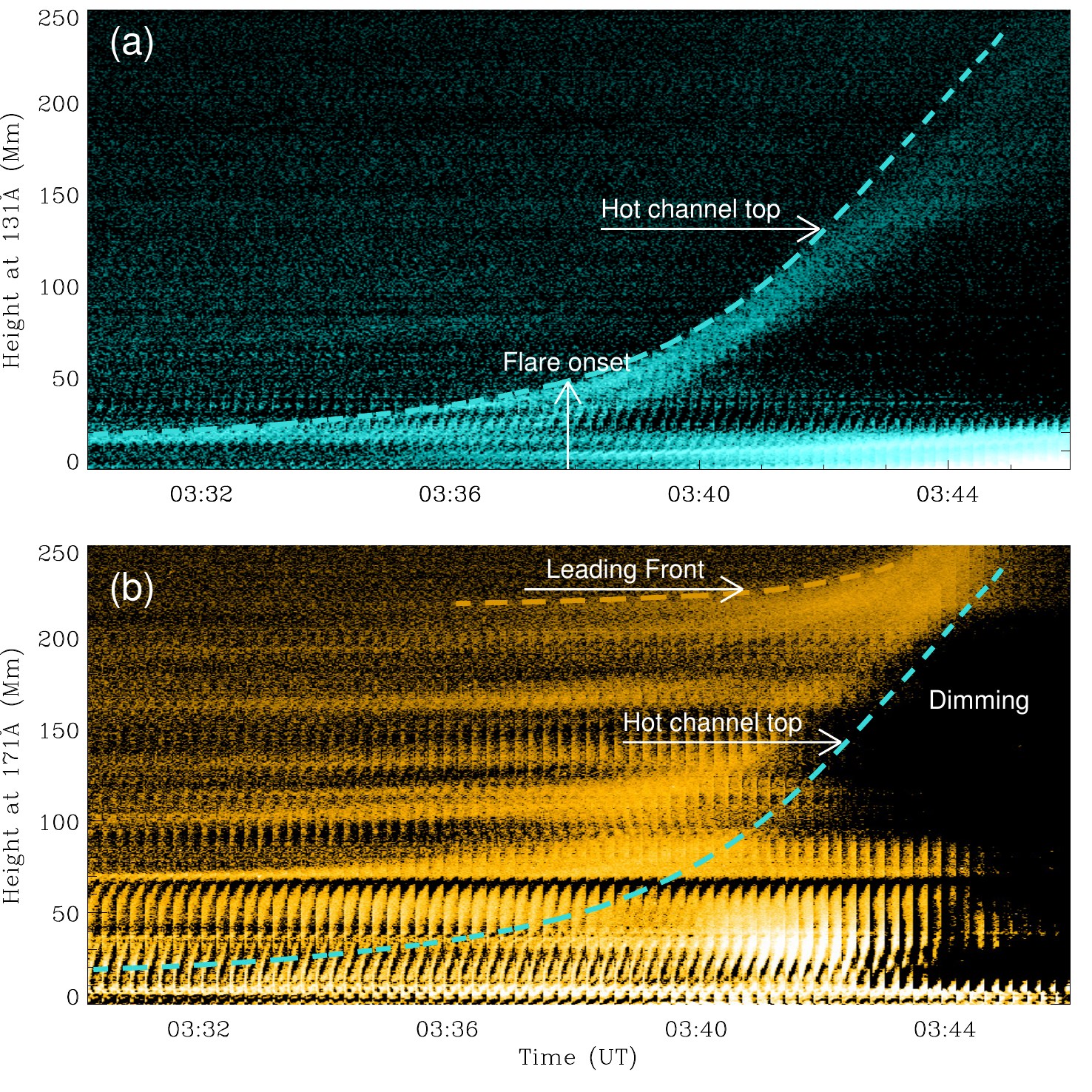}
               }
\vspace{0.0\textwidth}   
\caption{Temporal evolution of the brightness along the slice as shown in Figure \ref{f6}(b) for 131 {\AA} ($\sim$0.4 MK and 11 MK; upper) and 171 {\AA} ($\sim$0.6 MK; lower) base-difference images. The dashed line depicts the height evolution of the hot channel top. The vertical arrow indicates the onset of the flare rise phase.} \label{f8}
\end{figure}

\begin{figure} 
     \vspace{-0.1\textwidth}    
     \centerline{\hspace*{0.00\textwidth}
               \includegraphics[width=0.9 \textwidth,clip=]{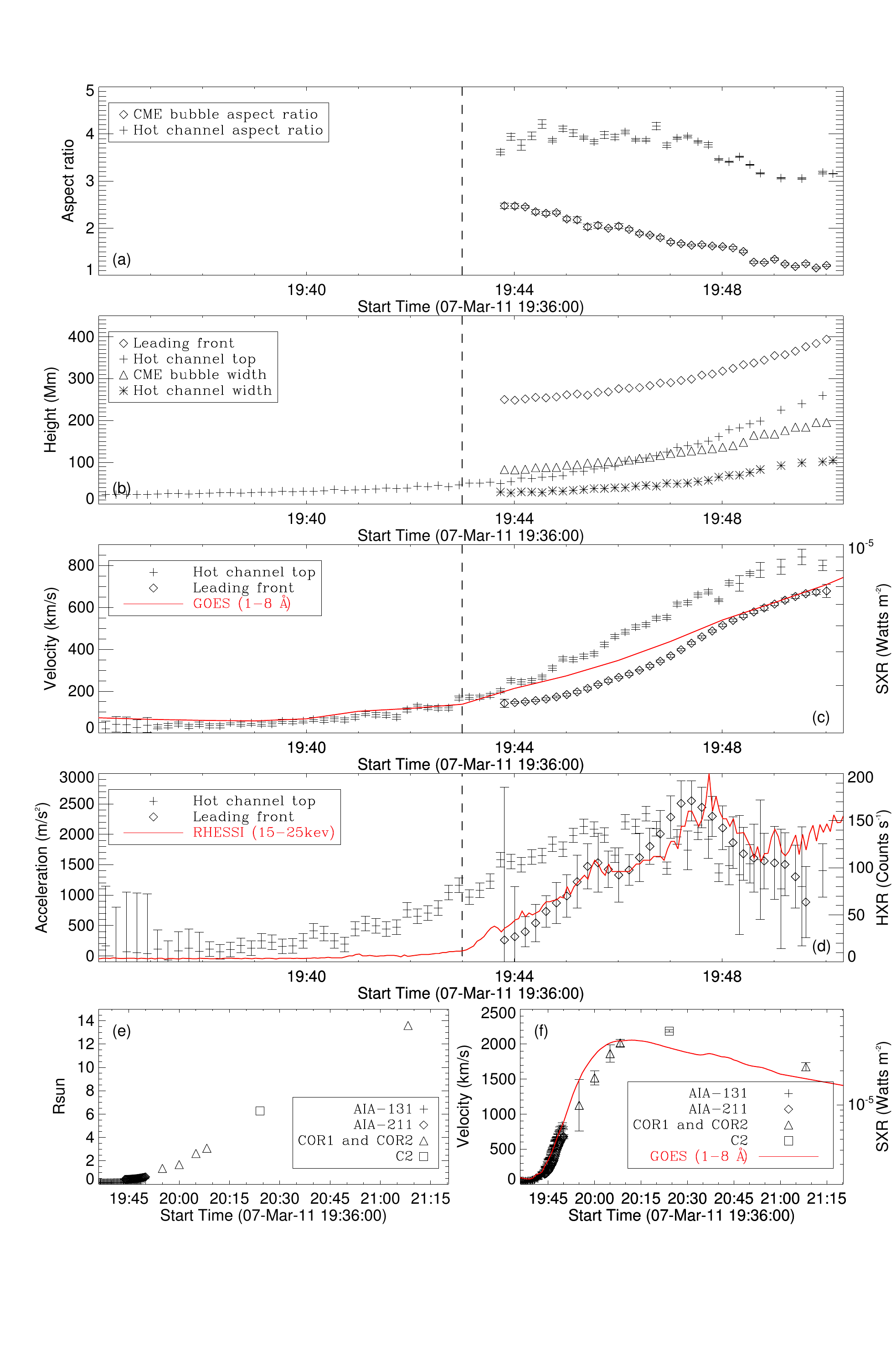}
               }
\vspace{-0.12\textwidth}   
\caption{Temporal evolution of the aspect ratios of the 2011 March 7 CME hot channel and bubble (a), the height and width of the CME hot channel, LF and bubble (b), the speed of the CME hot channel and LE (c), and the acceleration of the CME hot channel and LF (d). All of measurements in panels a-d are for the FOV of AIA. The vertical dashed lines indicate the onset of the flare rise phase. (e--f) Temporal evolution of the height and speed of the CME LF in a larger FOV. The GOES SXR 1--8 {\AA} flux of the associated flare is shown in panels c and f; the RHESSI 15--25 keV HXR flux is shown in panel d.}
\label{f9}
\end{figure}

\begin{figure} 
     \vspace{-0.0\textwidth}    
     \centerline{\hspace*{0.00\textwidth}
               \includegraphics[width=0.9 \textwidth,clip=]{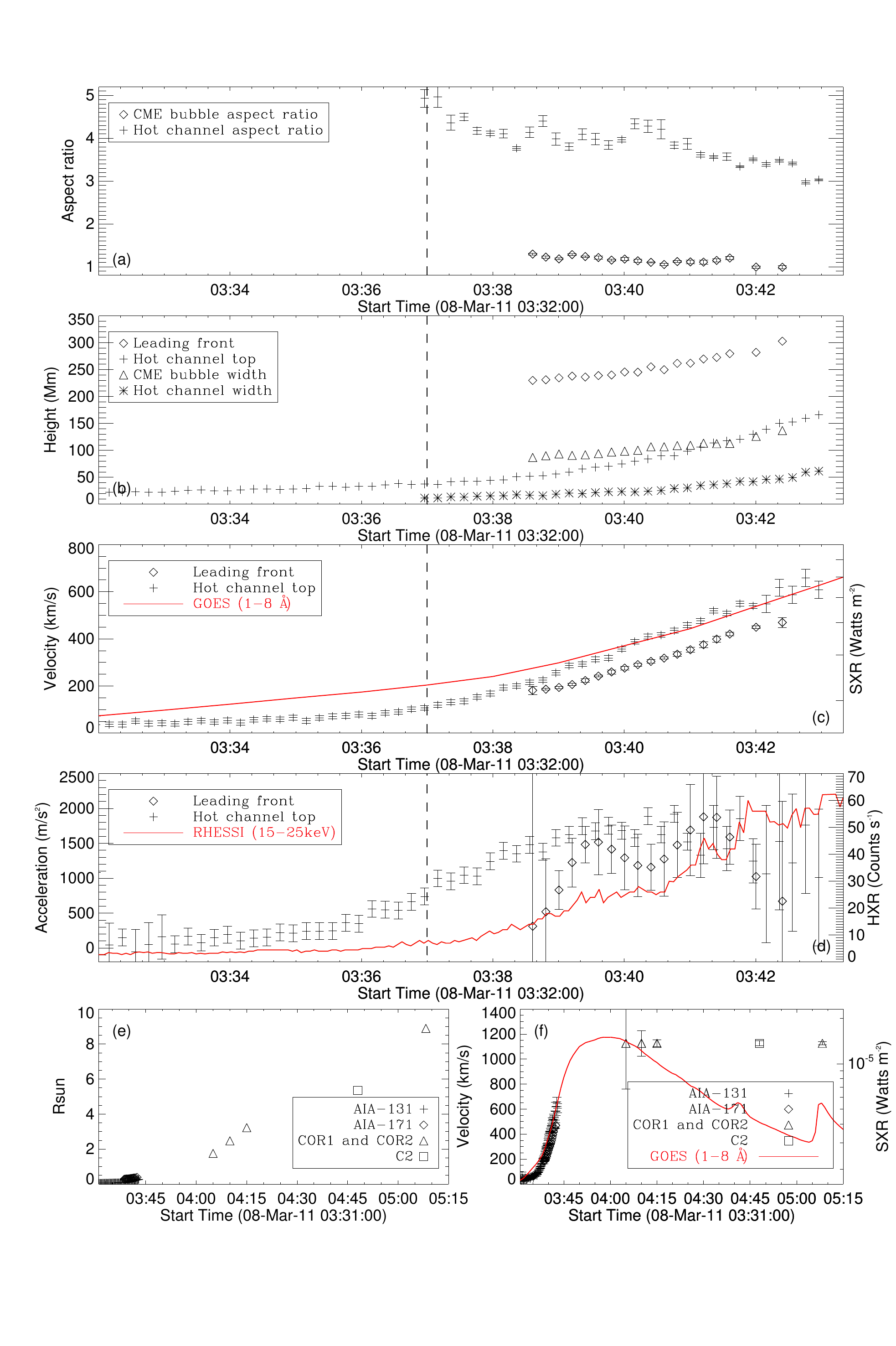}
               }
\vspace{-0.12\textwidth}   
\caption{Same as Figure 8 but for the 2011 March 08 CME.} \label{f10}
\end{figure}

\begin{figure} 
     \vspace{-0.0\textwidth}    
     \centerline{\hspace*{0.00\textwidth}
               \includegraphics[width=1.\textwidth,clip=]{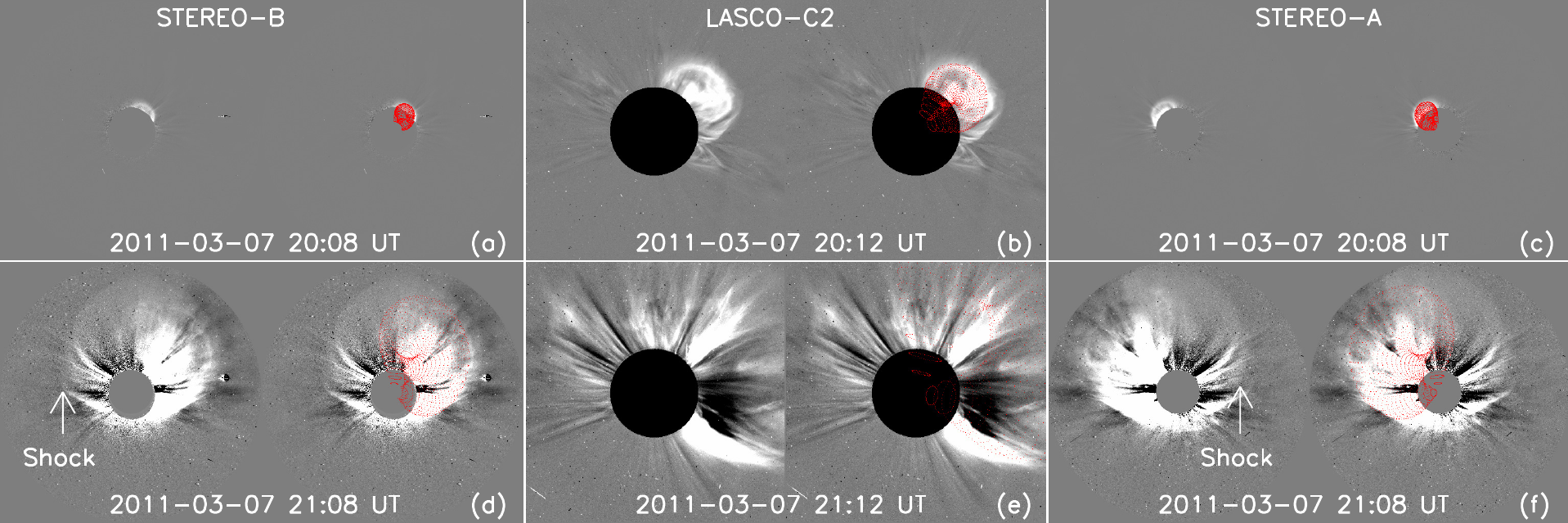}
               }
\vspace{0.0\textwidth}   
\caption{GCS flux rope (red lines) modeling of the 2011 March 07 CME.}
\label{f11}
\end{figure}

\begin{figure} 
     \vspace{-0.0\textwidth}    
     \centerline{\hspace*{0.00\textwidth}
               \includegraphics[width=1.\textwidth,clip=]{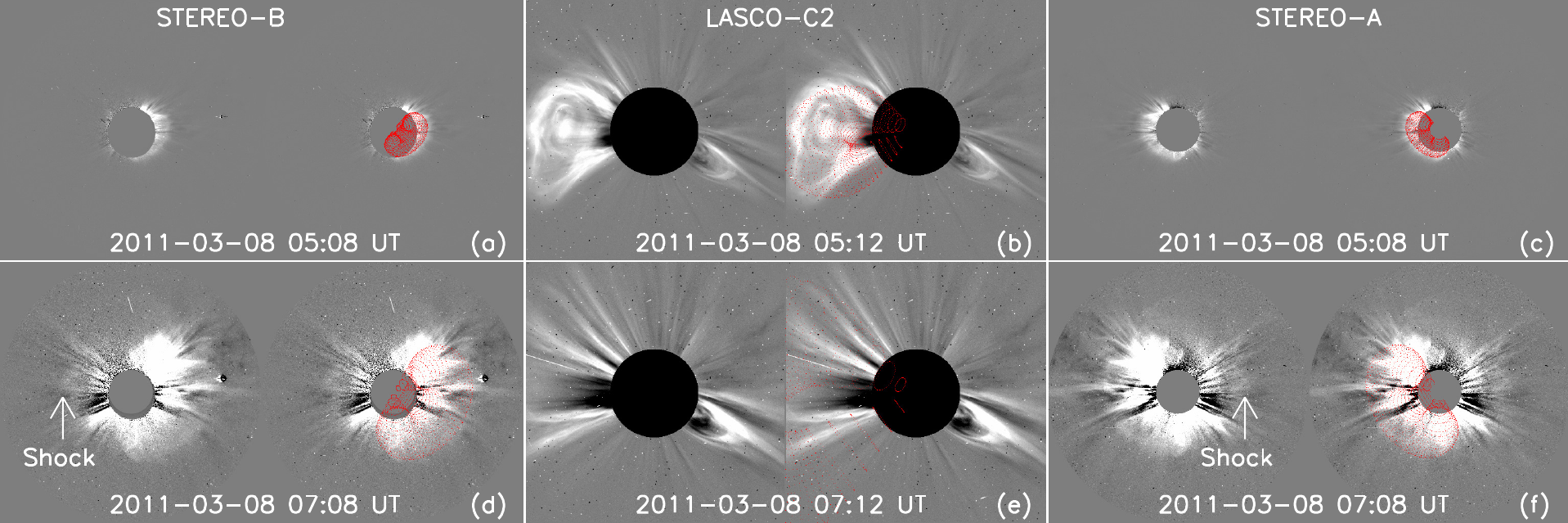}
               }
\vspace{0.0\textwidth}   
\caption{GCS flux rope (red lines) modeling of the 2011 March 08 CME.}
\label{f12}
\end{figure}

In the acceleration phase, the speed of the CME hot channel is larger than that of the CME LF, at least in the FOV of AIA. In a classic eruptive flare model \citep{carmichael64,sturrock66,hirayama74,kopp76}, when the explosive magnetic reconnection underneath the CME flux rope is triggered, the anchored overlying magnetic fields are quickly stretched upward by the erupting flux rope, and the magnetic field underneath is pushed together and forms a current sheet (CS). The reconnection in the CS converts more and more poloidal magnetic fluxes into the flux rope. The enhanced poloidal flux in the flux rope would essentially increase the upward Lorentz self-force and thus further accelerate the flux rope outward. This is similar to the flux injection model of \citet{chen96}, although his model is not reconnection driven but the flux comes from below the photosphere. In response to the fast escaping of the flux rope, a plasma inflow would form, which in turn makes more ambient magnetic fields reconnect in the CS. Thus, it is this positive feedback reconnection process that impulsively accelerates the CME flux rope. It is found that the hot channel studied here behaves just like the flux rope in the classic model. After the associated flare begins, the hot channel rapidly rises and expands through the positive feedback reconnection process into a full-fledged large-scale flux rope with the CS underneath and two footpoints still anchored at the ends of the main PIL. In the meantime, the flux rope compresses the overlying field and plasma that successively stack into the CME LF. Since the hot channel drives the formation of the CME LF, it is reasonable that the speed of the hot channel is larger than that of the LF in the early acceleration phase. Combining with the fact that the observed morphological evolution resembles that of the magnetic flux rope in theoretical models and 3D numerical simulations, we thus argue that the hot channel is possibly to be the magnetic flux rope of CMEs, which acts as a central engine to drive the CME formation and eruption \citep[see also,][]{zhang12}. However, it is worth noting that the flux rope top may catch up and merge with the CME LF quickly (e.g., the 2011 March 8 hot channel; Figure \ref{f8}(b)), making the LF sharper and brighter. The main flux rope body takes the form of the dark cavity in white-light coronagraphs. A detailed study of building the connection between the CME cavity in white-light observations to the hot channel and bubble in the EUV passbands is under the way.

In addition, we have studied the aspect ratio (height to width) of the hot channel and find that it tends to decrease during the impulsive acceleration phase. This means that the expansion speed of the hot channel is larger than its bulk propagation speed, which is defined as an overexpansion of the hot channel in this paper. The overexpansion possibly results from the magnetic reconnection in the acceleration phase, which rapidly transfers the ambient fields into the hot channel and causes the increase of the poloidal flux. An ideal MHD process can also result in the overexpansion of the flux rope-like hot channel. During the rise phase of the hot channel, the poloidal flux in the channel decreases because its twist does not change. In response to the decreased polodial flux, the channel will expand laterally \citep{pat10a}. In addition, the internal pressure (thermal plus magnetic) of the hot channel may be much larger than its external pressure, which can also drive the lateral expansion.

We believe that the lateral expansion of the hot channel also drives the formation of the CME bubble in the early acceleration phase. For the 2011 March 7 and 8 CMEs, it is found that the aspect ratio of the CME bubble gradually or slightly decreases after the beginning of the flare, indicating that the CME bubble also undergoes an overexpanding process like the hot channel \citep[also see,][]{pat10a,pat10b}. However, our observational results suggest that the overexpansion of the CME bubble may be the result of the overexpansion of the hot channel; when the small-scale hot channel (the nascent flux rope) is developing into the large-scale CME flux rope structure, it may result in the lateral expansion of surrounding magnetic loops, quickly forming the CME bubble and driving the bubble expansion in the low corona. Therefore, the overexpansion of the CME bubble is actually a manifestation of the dynamics of the hot channel.

\acknowledgements We thank the anonymous referee for her/his constructive comments that have significantly improved this manuscript. We are grateful to O. Olmedo, Q. R. Chen, and S. Patsourakos for valuable discussions. SDO is a mission of NASA's Living With a Star Program. X.C., and M.D.D. are supported by NSFC under grants 10673004, 10828306, and 10933003 and NKBRSF under grant 2011CB811402. X.C. is also supported by the scholarship granted by the China Scholarship Council (CSC) under file No. 2010619071. J.Z. is supported by NSF grant ATM-0748003, AGS-1156120 and NASA grant NNG05GG19G.


\bibliographystyle{apj}

\begin{thebibliography}{}
\expandafter\ifx\csname
natexlab\endcsname\relax\def\natexlab#1{#1}\fi

\bibitem[Amari et al.(2003)]{amari03} Amari, T., Luciani,
J.~F., Aly, J.~J., Mikic, Z., \& Linker, J.\ 2003, \apj, 585, 1073

\bibitem[Amari et al.(2000)]{amari00} Amari, T., Luciani,
J.~F., Mikic, Z., \& Linker, J.\ 2000, \apjl, 529, L49

\bibitem[Aulanier et al.(2010)]{aulanier10} Aulanier, G.,
T{\"o}r{\"o}k, T., D{\'e}moulin, P., \& DeLuca, E.~E.\ 2010, \apj, 708, 314

\bibitem[Burkepile et al.(2004)]{bur04} Burkepile, J.~T.,
Hundhausen, A.~J., Stanger, A.~L., St.~Cyr, O.~C., \& Seiden, J.~A.\
2004, Journal of Geophysical Research (Space Physics), 109, 3103

\bibitem[Brueckner et al.(1995)]{Brueckner95} Brueckner, G.~E., et
al.\ 1995, \solphys, 162, 357

\bibitem[Canfield et al.(1999)]{canfield99} Canfield, R.~C.,
Hudson, H.~S., \& McKenzie, D.~E.\ 1999, \grl, 26, 627

\bibitem[Canou
\& Amari(2010)]{canou10} Canou, A., \& Amari, T.\ 2010, \apj, 715, 1566

\bibitem[Canou et al.(2009)]{canou09} Canou, A., Amari, T.,
Bommier, V., Schmieder, B., Aulanier, G., \& Li, H.\ 2009, \apjl,
693, L27

\bibitem[Carmichael(1964)]{carmichael64} Carmichael, H.\ 1964, NASA
Special Publication, 50, 451


\bibitem[Chen(1996)]{chen96} Chen, J.\ 1996, \jgr, 101, 27499

\bibitem[Chen(2011)]{chen11} Chen, P.~F.\ 2011, Living Reviews
in Solar Physics, 8, 1

\bibitem[Cheng et al.(2010b)]{cheng10b} Cheng, X., Ding, M.~D.,
Guo, Y., Zhang, J., Jing, J., \& Wiegelmann, T.\ 2010b, \apjl, 716,
L68

\bibitem[Cheng et al.(2010a)]{cheng10a} Cheng, X., Ding, M.~D.,
\& Zhang, J.\ 2010a, \apj, 712, 1302

\bibitem[Cheng et al.(2011)]{cheng11} Cheng, X., Zhang, J., Liu, Y.,
\& Ding, M.~D.\ 2011, \apjl, 732, L25

\bibitem[Cheng et al.(2012a)]{cheng12a} Cheng, X., Zhang, J.,
Olmedo, O., et al.\ 2012a, \apjl, 745, L5

\bibitem[Cheng et al.(2012b)]{cheng12b} Cheng, X., Zhang, J., Saar, S.,
\& Ding, M.~D.\ 2012b, \apj, 761, 62

\bibitem[Dai et al.(2012)]{dai12} Dai, Y., Ding, M.~D., Chen,
P.~F., \& Zhang, J.\ 2012, \apj, 759, 55 


\bibitem[Downs et al.(2012)]{downs12} Downs, C., Roussev,
I.~I., van der Holst, B., Lugaz, N.,
\& Sokolov, I.~V.\ 2012, \apj, 750, 134

\bibitem[Fan(2010)]{fan10} Fan, Y.\ 2010, \apj, 719, 728

\bibitem[Gibson et al.(2006)]{gibson06} Gibson, S.~E., Foster,
D., Burkepile, J., de Toma, G., \& Stanger, A.\ 2006, \apj, 641, 590

\bibitem[Gosling et al.(1993)]{gosling93} Gosling, J. T. 1993, \jgr, 98, 18937

\bibitem[Green et al.(2007)]{green07} Green, L.~M., Kliem, B.,
T{\"o}r{\"o}k, T., van Driel-Gesztelyi, L.,
\& Attrill, G.~D.~R.\ 2007, \solphys, 246, 365

\bibitem[Green et
al.(2011)]{green11} Green, L.~M., Kliem, B., \& Wallace, A.~J.\ 2011, \aap, 526, A2

\bibitem[Guo et al.(2010)]{guo10} Guo, Y., Schmieder, B.,
D{\'e}moulin, P., Wiegelmann, T., Aulanier, G., T{\"o}r{\"o}k, T.,
\& Bommier, V.\ 2010, \apj, 714, 343

\bibitem[Hirayama(1974)]{hirayama74} Hirayama, T.\ 1974, \solphys,
34, 323

\bibitem[Howard et al.(2008)]{howard08} Howard, R. A., et al. 2008, \ssr, 136, 67

\bibitem[Illing
\& Hundhausen(1983)]{illing83} Illing, R.~M.~E., \& Hundhausen, A.~J.\ 1983, \jgr, 88, 10210

\bibitem[Jing et al.(2010)]{jing10} Jing, J., Yuan, Y.,
Wiegelmann, T., Xu, Y., Liu, R., \& Wang, H.\ 2010, \apjl, 719, L56

\bibitem[Kliem \& T{\"o}r{\"o}k(2006)]{kliem06} Kliem, B.,
 \& T{\"o}r{\"o}k, T.\ 2006, \prl, 96, 255002

\bibitem[Kopp
\& Pneuman(1976)]{kopp76} Kopp, R.~A., \& Pneuman, G.~W.\ 1976, \solphys, 50, 85

\bibitem[Lemen et al.(2011)]{lemen11} Lemen, J.~R., et al.\
2011, \solphys, 106

\bibitem[Li et al.(2010)]{li11} Li, T., Zhang, J., Zhao, H.,
\& Yang, S.\ 2010, \apj, 720, 144

\bibitem[Lin \& Forbes(2000)]{lin00} Lin, J., \& Forbes, T. G., 2000, \jgr, 105, 2375

\bibitem[Lin et al.(2002)]{lin02} Lin, R. P., et al. 2002, \solphys, 210, 3

\bibitem[Liu et al.(2010)]{liu10} Liu, R., Liu, C., Wang, S.,
Deng, N., \& Wang, H.\ 2010, \apjl, 725, L84


\bibitem[Liu et al.(2008)]{liuy08} Liu, Y., Luhmann, J.~G.,
Huttunen, K.~E.~J., Lin, R.~P., Bale, S.~D., Russell, C.~T.,
\& Galvin, A.~B.\ 2008, \apjl, 677, L133

\bibitem[Liu et al.(2007)]{liuy07} Liu, Y., Manchester, W.~B.,
IV, Kasper, J.~C., Richardson, J.~D.,
\& Belcher, J.~W.\ 2007, \apj, 665, 1439

\bibitem[Liu et al.(2010)]{liuy10}
Liu, Y., Thernisien, A., Luhmann, J.~G., Vourlidas, A., Davies,
J.~A., Lin, R.~P., \& Bale, S.~D.\ 2010, \apj, 722, 1762

\bibitem[McKenzie
\& Canfield(2008)]{mckenzie08} McKenzie, D.~E., \& Canfield, R.~C.\ 2008, \aap, 481, L65

\bibitem[Moore et al.(2001)]{moore01} Moore, R.~L., Sterling,
A.~C., Hudson, H.~S., \& Lemen, J.~R.\ 2001, \apj, 552, 833

\bibitem[O'Dwyer et
al.(2010)]{odwyer10} O'Dwyer, B., Del Zanna, G., Mason, H.~E., Weber, M.~A., \& Tripathi, D.\ 2010, \aap, 521, A21

\bibitem[Olmedo et al.(2012)]{olmedo12} Olmedo, O., Vourlidas,
A., Zhang, J., \& Cheng, X.\ 2012, \apj, 756, 143

\bibitem[Olmedo
\& Zhang(2010)]{olmedo10} Olmedo, O., \& Zhang, J.\ 2010, \apj, 718, 433

\bibitem[Patsourakos
\& Vourlidas(2009)]{pat09} Patsourakos, S., \& Vourlidas, A.\ 2009, \apjl, 700, L182

\bibitem[Patsourakos
\& Vourlidas(2012)]{pat12} Patsourakos, S., \& Vourlidas, A.\ 2012, \solphys, 93

\bibitem[Patsourakos et
al.(2010a)]{pat10a} Patsourakos, S., Vourlidas, A., \& Kliem, B.\ 2010a, \aap, 522, A100

\bibitem[Patsourakos et al.(2010b)]{pat10b} Patsourakos, S.,
Vourlidas, A., \& Stenborg, G.\ 2010b, \apjl, 724, L188

\bibitem[Poomvises et al.(2010)]{yod10} Poomvises, W., Zhang,
J., \& Olmedo, O.\ 2010, \apjl, 717, L159

\bibitem[Priest
\& Forbes(2002)]{priest02} Priest, E.~R., \& Forbes, T.~G.\ 2002, \aapr, 10, 313

\bibitem[Qiu et al.(2004)]{qiu04} Qiu, J., Wang, H., Cheng, C. Z., \& Gary, D. E. 2004, \apj, 604 900

\bibitem[Riley et al.(2008)]{riley08} Riley, P., Lionello, R.,
Miki{\'c}, Z., \& Linker, J.\ 2008, \apj, 672, 1221

\bibitem[Rust
\& Kumar(1996)]{rust96} Rust, D.~M., \& Kumar, A.\ 1996, \apjl, 464, L199

\bibitem[Schou et al.(2012)]{schou12} Schou, J., Scherrer,
P.~H., Bush, R.~I., et al.\ 2012, \solphys, 275, 229

\bibitem[Sturrock(1966)]{sturrock66} Sturrock, P.~A.\ 1966, \nat,
211, 695

\bibitem[Su et al.(2012)]{su12} Su, Y., Dennis, B.~R.,
Holman, G.~D., et al.\ 2012, \apjl, 746, L5

\bibitem[Temmer et al.(2008)]{temmer08} Temmer, M., Veronig, A. M.,
Vr\v{s}nak, B., Ryb\'{a}k, J., G\"{o}m\"{o}ry, P., Stoiser, S., \&
Mari\v{c}i\'{c}, D. 2008, \apj, 673, L95

\bibitem[Temmer et al.(2010)]{temmer10} Temmer, M., Veronig,
A.~M., Kontar, E.~P., Krucker, S., \& Vr{\v s}nak, B.\ 2010, \apj,
712, 1410

\bibitem[Thernisien et al.(2006)]{thernisien06} Thernisien,
A.~F.~R., Howard, R.~A., \& Vourlidas, A.\ 2006, \apj, 652, 763

\bibitem[Thernisien et al.(2009)]{thernisien09} Thernisien, A.,
Vourlidas, A., \& Howard, R.~A.\ 2009, \solphys, 256, 111

\bibitem[T{\"o}r{\"o}k
\& Kliem(2003)]{torok03} T{\"o}r{\"o}k, T., \& Kliem, B.\ 2003,
\aap, 406, 1043

\bibitem[T{\"o}r{\"o}k
\& Kliem(2005)]{torok05} T{\"o}r{\"o}k, T., \& Kliem, B.\ 2005,
\apjl, 630, L97

\bibitem[Vourlidas et al.(2012)]{vourlidas2012} Vourlidas, A.,
Syntelis, P., \& Tsinganos, K.\ 2012, \solphys, 27

\bibitem[Vr{\v s}nak et al.(2007)]{vrsnak07} Vr{\v s}nak, B.,
Mari{\v c}i{\'c}, D., Stanger, A.~L., Veronig, A.~M., Temmer, M.,
\& Ro{\v s}a, D.\ 2007, \solphys, 241, 85

\bibitem[Webb et al.(1994)]{webb94} Webb, D. F., Forbes, T. G., \&
Aurass, H. et al. 1994, \solphys, 153, 73

\bibitem[Williams et al.(2005)]{williams05} Williams, D.~R.,
T{\"o}r{\"o}k, T., D{\'e}moulin, P., van Driel-Gesztelyi, L.,
\& Kliem, B.\ 2005, \apjl, 628, L163

\bibitem[Zhang et al. (2012)]{zhang12} Zhang, J., Cheng, X., \& Ding, M. D. 2012, Nature Communications, 3, 747

\bibitem[Zhang \& Dere(2006)]{zhang06} Zhang, J., \& Dere, K. P. 2006, \apj, 649, 1100

\bibitem[Zhang et al.(2001)]{zhang01} Zhang, J., Dere, K. P., Howard, R. A., Kundu, M. R., \& White, S. M. 2001, \apj, 559, 452

\bibitem[Zhang et al.(2004)]{zhang04} Zhang, J., Dere, K. P., Howard, R. A. \& Vourlidas, A. 2004, \apj, 604, 420

\end{thebibliography}


\begin{table*}
\caption{Model and Positioning Parameters of CME flux ropes.}
\label{tb}
\begin{tabular}{cccccccc}
\\ \tableline \tableline
 Event         &  Time     & Lon ($\phi$)   & Lat ($\theta$)  & Tilt ($\gamma$) & H ($r$)      & Ratio ($\alpha$)  & Half-angle ($\kappa$)       \\
               &  (UT)     & (Deg)          &   (Deg)         & (Deg)           & (R$_\odot$)  &    (r/H)          &   (Deg)          \\
\hline
2011-03-07 CME & 20:08        &156.5              &29.6      &--77.1        &4.4        &  0.4     &  15.6         \\
               & 21:08        &156.5              &29.6      &--86.6        &15.2       &  0.4     &  40.2         \\
               \hline
2011-03-08 CME & 05:08        &30.2               &--10.6    &43.0          &6.7        &  0.3    &  32.7         \\
               & 06:08        &30.2               &--10.6    &45.8          &11.1       &  0.3    &  35.7         \\
               & 07:08        &30.2               &--10.6    &54.2          &15.3       &  0.3    &  35.7         \\
               \tableline
\end{tabular}
\end{table*}
\end{document}